# Multi-carrier transport in ZrTe$_5$ film[*]

Fangdong Tang(汤方栋)[1,2], Peipei Wang(王培培)[2], Peng Wang(王鹏)[2]，Yuan Gan(甘远)[2], Le Wang(王乐)[1†], Wei Zhang(张威)[1], and Liyuan Zhang(张立源)[2]

[1]*Department of Physics and Beijing Key Laboratory of Optoelectronic Functional Natual Materials & Micro-nano Devices, Renmin University of China, Beijing 100872, China*

[2]*Department of Physics, Southern University of Science and Technology, Shenzhen 518055, China*

The single layer of Zirconium pentatelluride (ZrTe$_5$) has been predicted to be a large-gap two-dimensional (2D) topological insulator, which has attracted particular attention in the topological phase transitions and potential device application. Here we investigated the transport properties in ZrTe$_5$ films with the dependence of thickness from a few nm to several hundred nm. We find that the temperature of the resistivity anomaly's peak (T$_p$) is inclining to increase as the thickness decreases, and around a critical thickness of ~40 nm, the dominating carriers in the films change from n-type to p-type. With comprehensive studying of the Shubnikov-de Hass (SdH) oscillations and Hall resistance at variable temperatures, we demonstrate the multi-carrier transport instinct in the thin films. We extract the carrier densities and mobilities of two majority carriers using the simplified two-carrier model. The electron carriers can be attributed to the Dirac band with a non-trivial Berry's phase $\pi$, while the hole carriers may originate from the surface chemical reaction or unintentional doping during the microfabrication process. It is necessary to encapsulate ZrTe$_5$ film in the inert or vacuum environment to make a substantial improvement in the device quality.

[*] Project supported by Guangdong Innovative and Entrepreneurial Research Team Program (No.2016ZT06D348) and Shenzhen Peacock Program (No: KQTD2016022619565991)
[†] Corresponding author. E-mail: le.wang@ruc.edu.cn

## 1. Introduction

The topological insulators have provided an exciting platform for exploring novel quantum phases and phenomena, such as the quantum spin Hall (QSH) effect[1-3], and the quantum anomalous Hall effect[4, 5], etc. ZrTe$_5$ is an orthorhombic layered material, which has been predicted as a 2D topological insulator in the monolayer [6]. The ZrTe$_5$ bulk system also displays the abundant electronic and exotic transport properties, and the topological nature is still unclear until now, i.e. strong topological insulator [7], weak topological insulator[8-10], and Dirac semimetal[11-13]. However, there are still very less experimental studies on ZrTe$_5$ thin films. And it is not fully understood of the conversion between 3D and 2D systems. Since the interlayer coupling in ZrTe$_5$ is as weak as graphite, it allows us to reduce sample's thickness to monolayer by mechanical exfoliation[14]. More recently, several groups have discovered that the ZrTe$_5$'s transport properties can be changed by the thickness and temperature, such as the Hall reversal[15, 16], the resistivity-anomaly temperature ($T_p$) shifting [17, 18], and the strong anisotropic transport behaviors in angular geometry devices[19]. The lastest angle-resolved photoemission spectroscopy (ARPES) and scanning-tunneling microscopy (STM) results provid strong evidences for the topological protected metallic state at step edge[9], which could host the quantum spin hall state. Altogether, these phenomena require further transport investigation of the interplay with the electrical field, magnetic field, temperature, and thickness, for the aim to clarify the still unknown topological nature and resistivity anomaly of the thin ZrTe$_5$ film.

In this work, we have systematically performed the transport measurements in ZrTe$_5$ thin films, with tuning thickness, temperature, electrical field and magnetic field. The shifting of $T_p$ and Hall resistance ($R_{xy}$) anomaly accompanied by the SdH oscillations are observed when reducing the thickness, which demonstrate the multi-carrier scenario in thin films. Using the simplified two-carrier model, we figure out the carrier density and mobility of each band. One is Dirac-like electron band with high mobility of ~20,000 $cm^2V^{-1}s^{-1}$ and low carrier density of ~$10^{16}$-$10^{17}$ $cm^{-3}$, hosting the nontrivial Berry's phase clarified by the Landau-Fan diagram analysis of the SdH oscillations[16]. The other is trivil hole-like band with low mobility of $10^2$~$10^3$ $cm^2V^{-1}s^{-1}$ and high carrier density of $10^{18}$~$10^{19}$ $cm^{-3}$. Noticing the decay of sample in ambient condition, this hole band may originate from the hole doping by the surface chemical reaction. The similar results have also been discussed in several other semimetals, such as $Cd_3As_2$[20], $WTe_2$[21], $HfTe_5$[22], $TaAs_2$[23], and graphite[24]. These observations provide a clue to understand the connection between 2D and 3D $ZrTe_5$, and to make a substantial improvement in the device quality.

## 2. Experimental methods

The single crystal of $ZrTe_5$ was grown by chemical vapor transport (CVT) method[25]. The $ZrTe_5$ thin films were prepared by standard mechanical exfoliation from a bulk crystal, then deposited onto the silicon substrates with 285 nm $SiO_2$. The samples were first identified using optical microscopy, and then immediately coated by a PMMA film in the glove box filled with argon gas. The exact thickness was finally measured using Atomic Force Microscopy (model Keysight 5500) after transport measurement. The devices were fabricated by traditional e-beam lithography process in Hall bar geometry. For the thick films, we used PDMS as media to transfer the film onto a pre-fabricated electrodes pattern[26]. The magneto-transport measurements were carried out in an Oxford TeslatronPT cryostat with variable temperatures from 1.5 K to 300 K and a magnetic field up to 14 T. The standard lock-in method with low frequency (17.777 Hz) was used to measure the longitudinal and transverse resistivity. The typical measurement current was 10 nA ~ 1 μA depending on the resistance of the samples. The back gate experiment was carried out using a Keithley 2400 source meter.

## 3. Results and discussion

### 3.1 Thickness-dependent transport.

Figure 1(a) displays the temperature-dependent resistivity of the bulk $ZrTe_5$. The

$T_p$ of the bulk sample is 140.1 K, which is a typical value for a ZrTe$_5$ crystal grown by CVT. The ARPES and transport results have revealed that $T_p$ is the cut-off point of the temperature-induced Lifshitz transition of the Fermi surface topology[27, 28]. The relationship between $T_p$ and thickness is summarized in Figure 1(c), while the temperature-dependent resistivity curves of some thin films are plotted in Figure 1(b) -- the $T_p$ first decreases concerning the thickness down to 40 nm then increases when continuing to reduce the thickness[17]. Here it is necessary to point out that although the whole tendency indicates the thickness-tuned effect of $T_p$ in figure 1(c), the $T_p$s of thin films (< 40 nm) vary dramatically from 140 K to above 300 K. It is unadvised to draw the conclusion that a highly thickness-tuned transition of the band topology occurs in ZrTe$_5$ films.

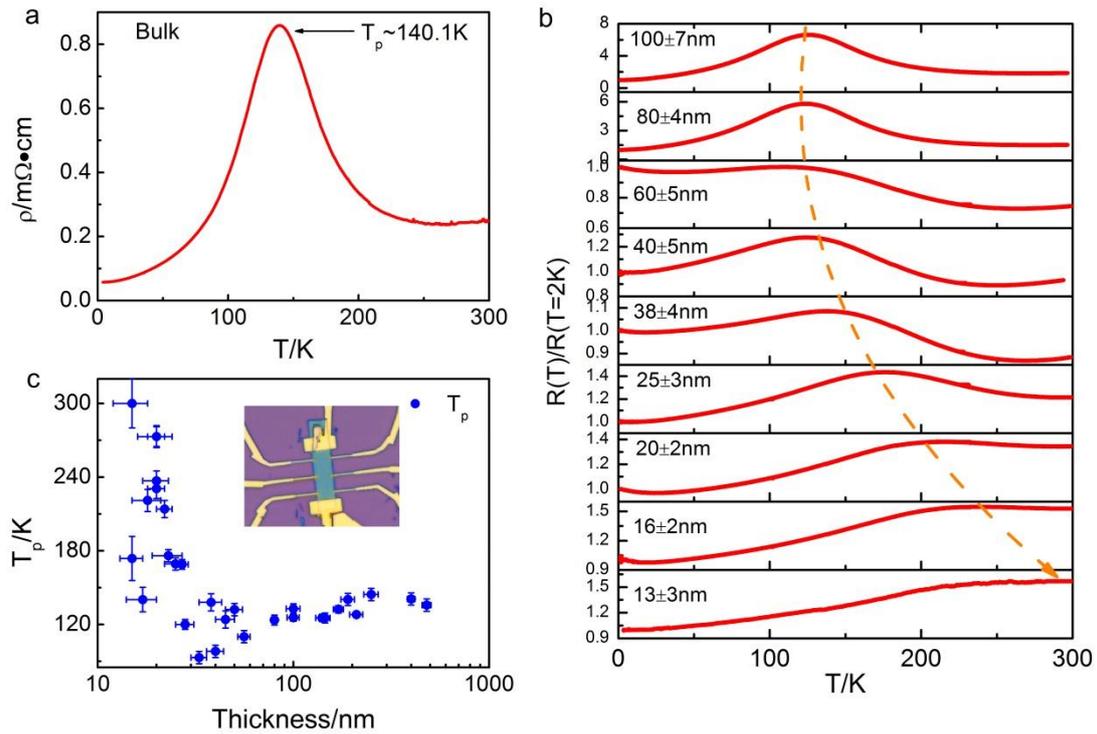

**Fig. 1.** Thickness dependence of resistance anomaly temperature in ZrTe$_5$ films. (a) The temperature dependence of resistivity of bulk ZrTe$_5$. The resistance anomaly temperature ($T_p$) is around 140.1 K. (b) Some selected R-T curves of thickness from 100 nm to 13 nm. (c) Systematic statics of thickness-dependent $T_p$ transition. The whole tendency is first decreasing then increasing with largely sample dependence, indicating the extreme sensitivity of thin samples.

To better understand the thickness-dependent transport property, the Hall resistance of these samples with different thickness is measured at 1.5 K. As shown in Figure 2(a), the slopes of Hall resistance $R_{xy}$ in thick samples (> 100 nm) are negative in the whole magnetic field region (14 T) and exhibit nonlinear behavior at high field. This nonlinear curve indicates the existence of more than one electron band participating in the transport. For thinner samples (33 nm to 100 nm), the Hall sign is negative at the low field but turns to positive at high field. This "Z" shaped Hall sign reversal implies the participation of hole-type band[20]. When the thickness is less than 33 nm, the Hall anomaly disappears and the slopes of Hall resistance become positive, indicating that the hole-type carriers dominate the transport. This Hall resistance transition is in well consistent with the change of $T_p$ in Figure 1. However, as shown in Figure 2(b), this regular behavior breaks in these samples whose thickness is located between 26 nm and 45 nm. More evidently, a clear difference of the magneto-resistance (MR), as well as the Hall resistance, is observed in these samples in one-week interval. Figure 2(c) gives the comparing results for an 80 nm thick sample. Even though the samples are stored in an argon protected glove box to minimize the surface oxidation, the MR becomes ~ 40.4% of the initial measured result, and the slope of Hall resistance exhibits larger hole doping of ~ 131% than the initial measured value.

This significant deviation pushes us to consider the effect of decay during the sample preparation. The decay takes place at the surface, where the activated atoms emerge after exfoliated from a large bulk crystal. The activated atoms can easily react with the oxygen, water, and organic solvent during fabrication, leading to the giant hole doping at the surface[29, 30]. In thick samples (> 100 nm), this effect of surface doping is limited due to the small surface-to-volume ratio. The transport is dominated by the electron carriers in the bulk $ZrTe_5$, giving rise to the negative slopes of Hall resistance. On the contrary, this hole doping at surface becomes strong enough in thin films, resulting in the movement of the Fermi level. For samples whose thickness is less than 33 nm (see figure 2(a)), the heavy surface hole-doping carriers dominate the transport, leading to the positive Hall coefficient. For samples whose thickness is located between 33 nm and 100 nm, the character of transport is the competitive result of the electron carriers and the surface hole-doping carriers. The level of surface hole doping will influence the transport measurement results, which is certified by the time-interval measurement of MR and Hall resistance as shown in figure 2(c).

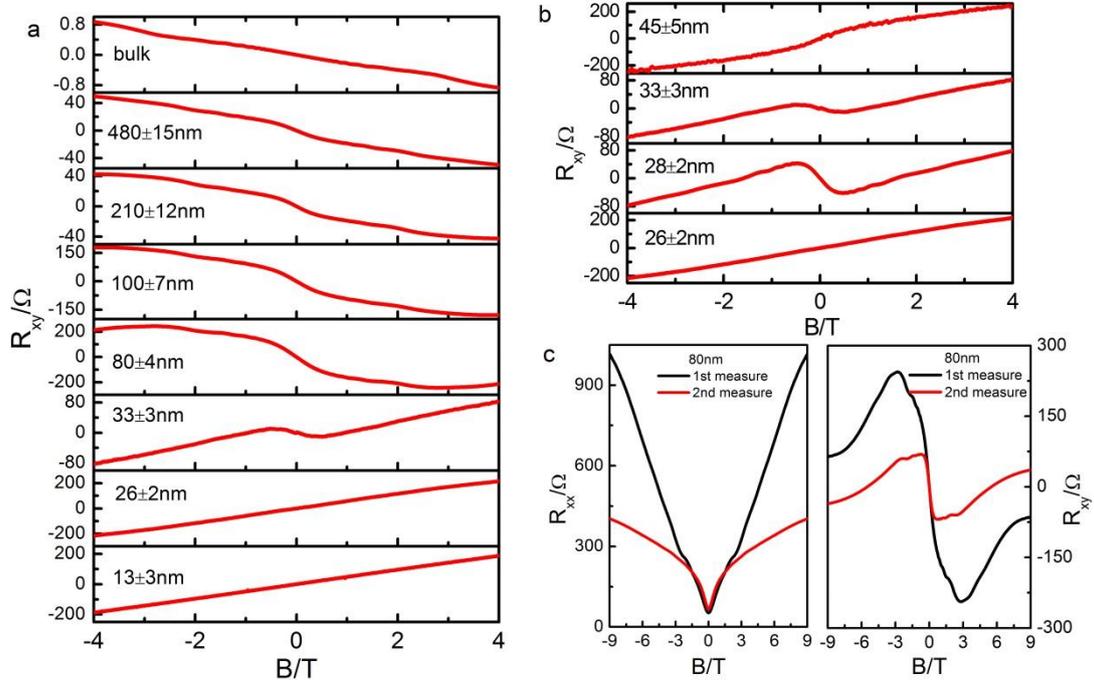

**Fig. 2.** (a) Some selected Hall resistance of different thickness in ZrTe$_5$ films at 1.5 K. It displays a "suspected" thickness tuned band transition from electron type to hole type when reducing the thickness. The Hall anomaly indicates the multi-carrier transport. (b) Some "irregular" thickness-dependent Hall resistance, largely depending on the sample's quality. (c) The comparison of magneto-resistance and Hall resistance of the same sample measured in one-week interval, and the sample is stored in the glove box with an argon atmosphere. Other samples also display the similar decay behavior.

### 3.2 Magnetic field-dependent transport.

Figure 3(a) and Figure 3(b) are the MR and Hall resistance of a 28 nm thick sample at temperature ranging from 1.5 K to 200 K, respectively. The MR shows a "V" shaped curve in low field accompanied by obvious SdH oscillations below 4 T, and shows a linear behavior above 4 T. The Hall resistance displays a "Z" shaped anomalous curve, and changes from negative to positive accompanied by SdH oscillations. This behavior is similar to that in Bi$_2$Te$_3$[31], which is a hallmark of the multi-carrier transport in most instances. The electron-to-hole transition disappears between 100 K ~140 K, corresponding to the T$_p$ (117 K) of this sample.

To further understand the SdH oscillations, the Landau plot and effective cyclotron mass fitting are performed. According to the Lifshitz-Kosevich formula[16, 32],

$$\Delta R \propto R_T R_D R_S cos2\pi \left(\frac{B_F}{B} + \gamma\right) \qquad (1)$$

Where $R_T$, $R_D$, and $R_S$ are three reducing factors accounting for the phase smearing effect of temperature, scattering, and spin splitting; $\gamma$ is a phase factor that implies the band singularity, from which Berry's phase of the system is obtained; $B_F$ is the slope of the Landau fan diagram; $B$ is the magnetic field. According to the Lifshitz-Onsage quantization rule: $B_F/B=N+1/2-\Phi_B/2\pi-\delta=N+\gamma$, where $N$ is the Landau level index, $\Phi_B$ is the Berry's phase, and $\delta$ is an additional phase shift which takes 0 for 2D or $\pm 1/8$ for 3D, $\gamma$ is the intercept in the Landau plot. Here, the peak values are chosen to be the integer indexes and the valleys are chosen to be the half-integer indexes. For Dirac materials, the $|\gamma|$ between 0 and 1/8 implies a non-trivial Berry's phase $\pi$ and around 1/2 indicates the trivial Berry's phase.

Figure 3(c) displays the oscillating amplitude of the resistance after subtracting the high-temperature background, in which the Zeeman splitting can also be observed[16]. Figure 3(d) displays the Landau-Fan diagram of samples in different thickness. All the intercepts are pointing to nearly zero, indicating the nontrivial Berry's phase, which proves to be the strong characteristic of Dirac band[32]. Furthermore, the slopes ($B_F$), which are directly proportional to the 2D carrier density, decrease when reducing the thickness, implying the lowering of the Fermi level due to the hole doping. The effective cyclotron mass m* ~ 0.057m$_e$, fitted by the data in the inset of figure 3(d), is at the same order of magnitude as the bulk material ~0.03m$_e$. The 3D carrier density of this band is roughly calculated and gives the value of $n_{3D}^{SdH} \sim (n_{2D}^{SdH})^{\frac{3}{2}} \sim 5.28 \times 10^{16}\ cm^{-3}$, using $g=2$ (the spin degeneracy of Landau level) and $B_F$ ~ 2.92 T. The carrier mobility can be estimated from $\mu_{SdH} \cdot B_{on} \sim 1$, leading to the value of 18,720 cm$^2$V$^{-1}$s$^{-1}$, where $B_{on}$ is the onset magnetic field of the SdH oscillations. From the above simplified analysis, the key parameters of this Dirac band are obtained.

We find that the nonlinear property in Hall resistance is coincident with the multi-carrier transport behavior. For simplification, the carriers are divided into two parts -- one is the Dirac band, and the other is regarded as an equivalent hole band with the same scattering time. Thus the standard two-carrier model is applied as[20, 24],

$$\sigma_{xx}^{total} = \frac{\rho_1}{\rho_1^2 + (R_1 B)^2} + \frac{\rho_2}{\rho_2^2 + (R_2 B)^2}$$
$$\sigma_{xy}^{total} = -\frac{R_1 B}{\rho_1^2 + (R_1 B)^2} - \frac{R_2 B}{\rho_2^2 + (R_2 B)^2} \quad (2)$$

Each band has two parameters: resistivity $\rho_i$ and Hall coefficient $R_i = 1/n_i q_i$, where $q_i = \pm e$ (i=1, 2) is the charge of the carrier. It is necessary to fit the $\sigma_{xx}$ and $\sigma_{xy}$ simultaneously by adjusting the four parameters independently until the differences between the fitting curves and the experimental data are minimized. We can also calculate that the Hall reverses its sign at a critical magnetic field $B_c = \sqrt{\frac{n_1 \mu_1^2 - n_2 \mu_2^2}{\mu_1^2 \mu_2^2 (n_2 - n_1)}}$, where $\mu_1$ represents the high mobility Dirac band and $\mu_2$ is the hole band. In this case $\mu_1 > \mu_2$, the decrease of electron concentration with increasing temperature will cause the decrease of $B_c$, in consistent with Figure 3(b). The fitting results are illustrated in figure 3(e) at representative 1.5 K, 60 K and 200 K. From the best fitting, the mobility and carrier density of the Dirac band at 1.5 K is 32,000 cm$^2$V$^{-1}$s$^{-1}$ and $5.90 \times 10^{16}$ cm$^{-3}$, in well consistent with the band parameters obtained from SdH oscillations. The mobility and carrier density of the hole band is 1,018 cm$^2$V$^{-1}$s$^{-1}$ and $2.25 \times 10^{18}$ cm$^{-3}$. In thinner samples, the hole mobility is even lowered to ~10$^2$ cm$^2$V$^{-1}$s$^{-1}$ and the carrier density increases to ~10$^{19}$ cm$^{-3}$. It is necessary to note that the two-carrier model is a classical model, which can't be used in quantum transport, with the "V" shaped MR and the SdH oscillations. However, the good description of the transport behavior here demonstrates the reasonability of the two-carrier model to some degree.

To reveal the two-carrier transport in the ZrTe$_5$ thin films, we use the Kohler's plots[33, 34] with $\frac{\Delta R_{xx}(B)}{R_{xx}(0)} \sim (\frac{B}{R_{xx}(0)})^2$ for the classical $B^2$ dependence of MR, as shown in Figure 3(f). If there is one type of carrier with the same scattering time at the Fermi surface everywhere, the temperature-dependent Kohler plot of the MR curve could overlap each other. At a temperature above 100 K, the collapsed curves indicate the hole-dominating transport, although there are two types of holes. And in low temperature, because of the increased Dirac electrons with high mobility, the curves separate from each other.

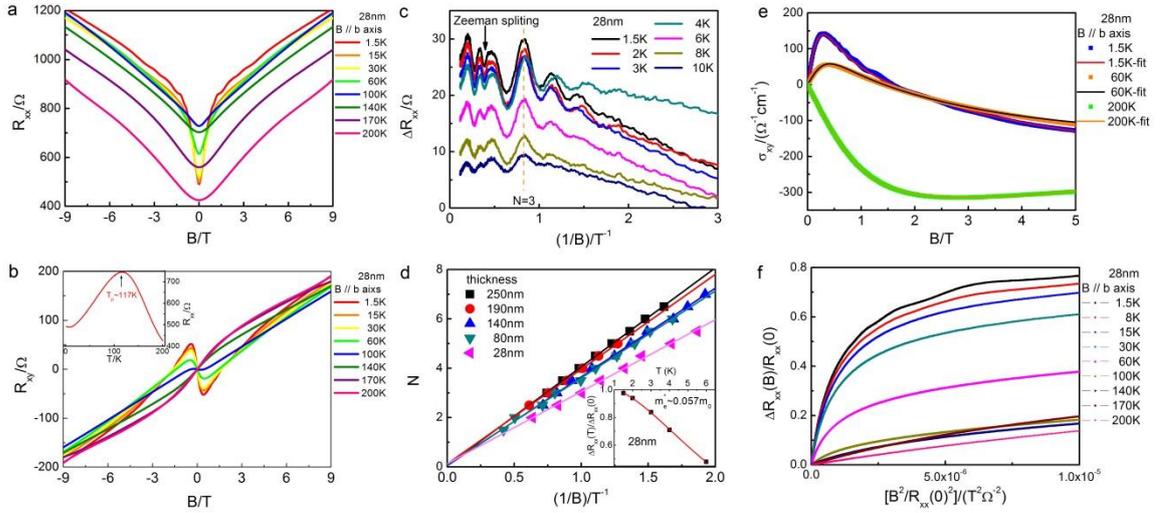

**Fig. 3.** Temperature-dependent magneto-transport of the 28 nm $ZrTe_5$ film. (a)(b) The longitudinal and transverse magneto-resistance of $ZrTe_5$ from 1.5 K to 200 K. The well-defined SdH oscillations are observed in the low field (< 4 T), and the Hall anomaly indicates two type carriers. (c) The amplitude of SdH oscillations at different temperatures after subtracting a high temperature background of 15 K, and the Zeeman splitting implies the well resolved Landau levels. (d) The Landau-Fan diagram of $ZrTe_5$ films in different thickness. The decrease in slope indicates the reduction in electron density. The zero intercepts demonstrate the nontrivial Berry's phase. The insert is the fit of effective cyclotron mass ($0.057m_e$) in ac plane. (e) The two-carrier model fits of Hall conductivity at 1.5 K, 60 K and 200 K respectively. (f) The Kohler's plot of MR at different temperature, indicating the hole dominating transport above 100 K and multi-carrier transport at low temperature.

**3.3 Field-effect transport.**

The gate-dependent longitudinal conductivity of thin films without the external magnetic field was also investigated. For thicker samples (> 30 nm), the gate tuned doping is limited because of the Thomas-Fermi dielectric screening[35, 36]. The gate voltage can only modulate about 5% ~ 20% times of original resistance without doping. For thin samples, the gate effect becomes larger, and most of the samples show p-type curves instead of ambipolar behavior. In thin films with less decay, it is possible to observe the ambipolar behavior. Figure 4(a) illustrates the transfer curves of a 15 nm thick sample from 3.6 K to 270 K. The insert shows the typical gate-tuned longitudinal conductivity at 3.6 K, in which a minimum of conductivity ($V_D$) at $V_g$ = 60 V is observed. The minimum point indicates the transition from the p-type to n-type. It should be noted that this transition point is different from the neutral point in semiconductors or graphene[37, 38], and we believe it is a comprehensive result of the contributions of the two carrier densities and mobilities, which is exceptionally evident in the electron-hole asymmetric system. This minimum point shifts toward positive gate voltage with increasing temperature, and disappears above 70 K continued by the whole p-type behavior. This transform may be caused by the temperature-induced Fermi level shifting down from conducting band to valence band.

The field-effect mobility $\mu_{FE}$ in the linear region of the transfer characteristics can be extracted by the following formula[39],

$$\mu_{FE} = \frac{\partial \sigma_{xx}}{\partial V_g} \frac{dt}{\varepsilon_0 \varepsilon_{ox}} \quad (3)$$

where $d$ and $t$ are the thickness of the film and the SiO$_2$ dielectric layer, respectively, $\varepsilon_{ox} = 3.9$ is the dielectric constant of SiO$_2$. By using the classical Drude model $\sigma_{xx} = ne\mu_{FE}$, the carrier density of the dominating p-type carrier can be roughly estimated, as shown in Figure 4(b). This estimation is reasonable because the hole carrier density is 2~3 orders larger than the electron carrier density, deduced by the two-carrier fitting in the previous section. When the temperature changes from 270 K to 3.6 K, the hole mobility increases from 192 cm$^2$V$^{-1}$s$^{-1}$ to 1,330 cm$^2$V$^{-1}$s$^{-1}$, and the carrier density decreases from 2.3×10$^{19}$ cm$^{-3}$ to 4.1×10$^{18}$ cm$^{-3}$. These two parameters are in well consistent to the results of two-carrier model fitting.

Figure 4(c) displays the temperature-dependent longitudinal resistivity under different applied gate voltages, extracted from the transfer curves in Figure 4(a). The resistivity anomaly points are summarized in Figure 4(d), in which the $T_p$ obviously changes from above 270 K down to 120 K when increasing the electron doping. This result provides a good explanation for the dramatical variation of the $T_p$s in thin films shown in Figure 1(c). When the sample gets thinner, the surface-to-volume ratio becomes larger, and it will easily get more hole doping if increasing the exposing time. Samples of the similar thickness may suffer heavier or lighter doping, which makes the $T_p$ changeable.

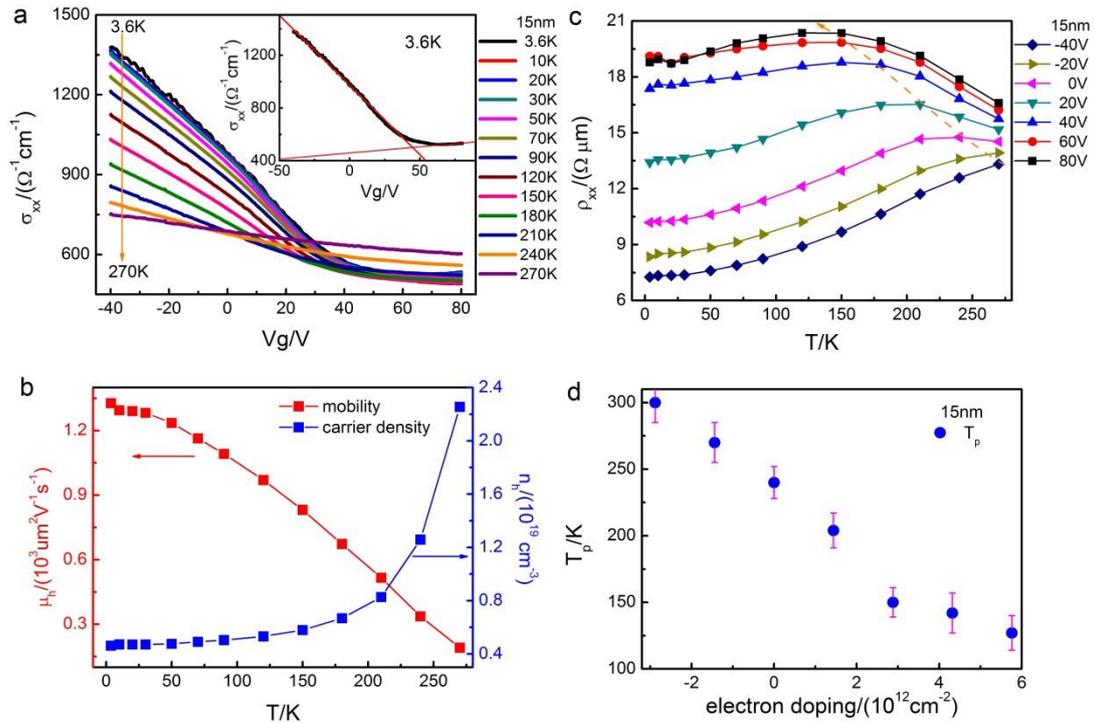

**Fig. 4.** (a) The gate voltage modulated longitudinal conductivity of the 15 nm $ZrTe_5$ films at different temperatures. It shows a peak around 60 V at 3.6 K. (b) The field effect mobility of the dominated hole type carriers and the carrier density roughly estimated by the Drude model. (c) The corresponding temperature-dependent longitudinal resistivity at different voltages displays the movement of $T_p$. (d) $T_p$ moves to lower temperature when increasing the electrons doping, indicating the easy modulation by the external doping level.

**3.4 Magneto-resistance with gating effect.**

Figure 5(a) shows the gate-tunable MR behaviors of the 80 nm thick sample at 1.5 K. The magnetic field is parallel to the b axis. The "V" shaped MR and the "Z" shaped Hall resistance under different gate voltages demonstrate the two-carrier transport character. As shown in Figure 5(b), all the Landau-plotting curves are overlapped in the same fitting line. Although different gate voltages are applied, the same Landau levels of the SdH oscillations still locate in the same magnetic field. We believe that most of the doped carriers go into the p-type band of the surface, instead of doping to the Dirac band. This result corresponds to the fitting result of the two-carrier model, and indicates that the carrier density and mobility of the Dirac band hardly change, while the hole-type carrier density decreases when applying larger gate voltage.

Finally, we study the gate effect of the $ZrTe_5$ devices at the different magnetic field[19]. As shown in Figure 5(c) and Figure 5(d), the transfer curves of a 31 nm thick device show the monotonous p-type behavior in all the field ranges, indicating the hole dominating transport. The transfer mobility is about 690 $cm^2V^{-1}s^{-1}$, clarified by the fitting value of 770 $cm^2V^{-1}s^{-1}$ from the Hall measurement, as shown in the insert figure. The small increase of the Hall resistance implies the decrease of the population of the hole (seen in Figure 5(d)). No SdH oscillations are observed in this thin sample. The reason may be the low mobility and high carrier density in the heavy hole-doping sample.

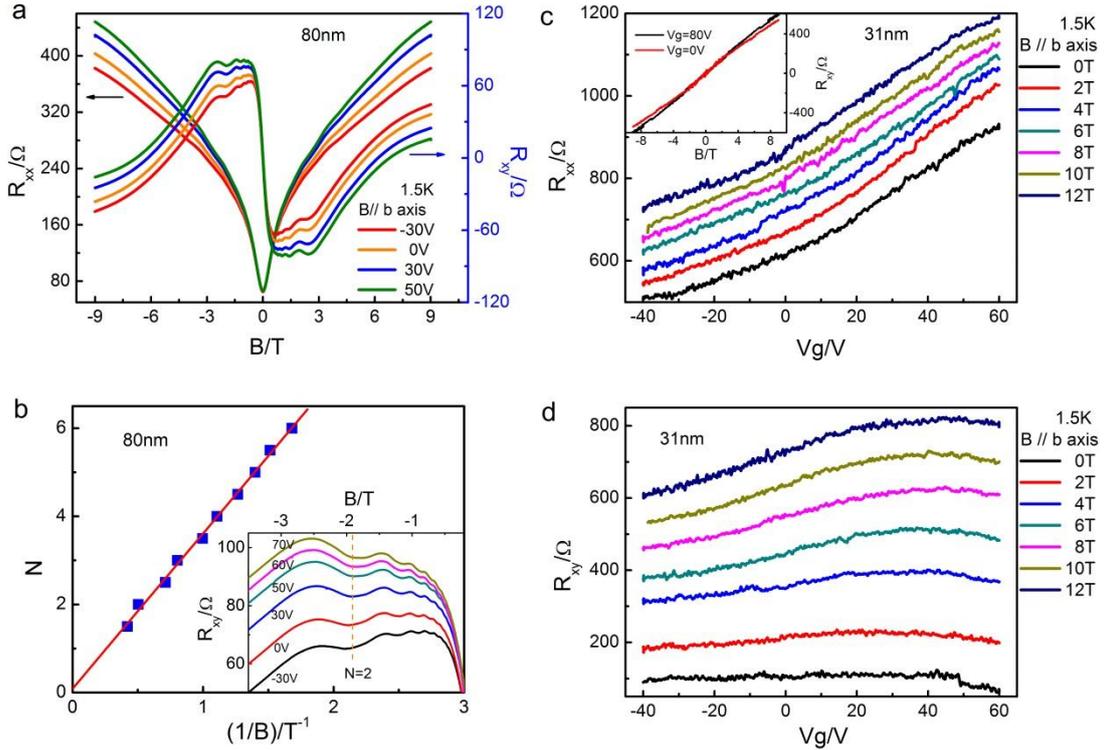

**Fig. 5.** The gate voltage modulated magnetotransport at 1.5 K. (a) The MR and Hall resistance of the 80 nm sample. (b) The Landau plot of the SdH oscillations, indicating the nontrivial Berry's phase. The insert is the Landau levels at different gate voltage. (c)(d) The p-type gate effect of MR and Hall resistance at the different magnetic field of the 31 nm sample.

## 4. Conclusions

In summary, we have systematically performed the low temperature magnetotransport measurement on the thickness-dependence, temperature–tuning, and gate-modulation in ZrTe$_5$ films. We find that T$_p$ is keeping increasing trend when the ZrTe$_5$ get thinner. However, the T$_p$s are irregular and exhibit large sample dependence, especially below 40 nm. Combining the SdH oscillations and Hall measurement, we extract the carrier densities and mobilities by two-carrier model fitting: one electron band hosts high mobility of ~20,000 cm$^2$V$^{-1}$s$^{-1}$ and low carrier density of ~10$^{16}$-10$^{17}$ cm$^{-3}$; the other hole band has low mobility of 10$^2$~10$^3$ cm$^2$V$^{-1}$s$^{-1}$ and high carrier density of 10$^{18}$~10$^{19}$ cm$^{-3}$. Considering the effect of decay on the sample surface, we propose that the multi-carrier tronsport property is induced by the coexistence of the Dirac-like electron band with nontrivial Berry's phase, and the trivial hole-like band from unintended surface chemical reaction. We anticipate that these results will be helpful to understand the conversion between bulk system to thin film. The more strict fabricating process will be needed by using the uniform insulating substrate (i.e., Hexoganol Boron-Nitride) as the capping layer, which may direct a reliable way to the study of the QSH effect and the topological phase transitions in the ZrTe$_5$ film.